\documentclass[12pt]{article}
\usepackage{graphicx,amsfonts,amsmath,amssymb,amsthm,mathrsfs,url,hyperref,tikz}

\title{The Problem of Boltzmann Brains and How Bohmian Mechanics Helps Solve It}

\author{
Roderich Tumulka\footnote{Mathematisches Institut,
     Eberhard-Karls-Universit\"at, Auf der Morgenstelle 10, 72076
     T\"ubingen, Germany. E-mail:
     roderich.tumulka@uni-tuebingen.de}
}
\date{December 5, 2018}

\renewcommand{\Im}{\mathrm{Im}}
\newcommand{\RRR}{\mathbb{R}}

\newcommand{\vQ}{\boldsymbol{Q}}

\newcommand{\be}{\begin{equation}}
\newcommand{\ee}{\end{equation}}
\newtheorem{thm}{Theorem}

\begin{document}
\maketitle
\begin{abstract}
Most versions of classical physics imply that if the 4-volume of the entire space-time is infinite or at least extremely large, then random fluctuations in the matter will by coincidence create copies of us in remote places, so called ``Boltzmann brains.'' That is a problem because it leads to the wrong prediction that \emph{we} should be Boltzmann brains. The question arises, how can any theory avoid making this wrong prediction? In quantum physics, it turns out that the discussion requires a formulation of quantum theory that is more precise than the orthodox interpretation. Using Bohmian mechanics for this purpose, we point out a possible solution to the problem based on the phenomenon of ``freezing'' of configurations.

\medskip

\end{abstract}

\section{The problem of Boltzmann brains}

A ``Boltzmann brain'' \cite{Alb02,BCP15} is this: Let $M$ be the present macro-state of your brain. For a classical gas in thermal equilibrium, it has probability 1 that after sufficient waiting time, some atoms will ``by coincidence'' (``by fluctuation,''  ``by ergodicity'') come together in such a way as to form a subsystem in a micro-state belonging to $M$. That is, this brain comes into existence not by childhood and evolution of life forms, but by coincidence; this brain has memories (duplicates of your present memories), but they are false memories: the events described in the memories never happened to this brain!
Boltzmann brains are, of course, very unlikely. They are fluctuations in which entropy spontaneously decreases. But they will happen if the waiting time is long enough, and they will happen more frequently if the system is larger (bigger volume, higher number of particles).

The problem with Boltzmann brains is this: If the universe continues to exist forever, and if it reaches universal thermal equilibrium at some point, then the overwhelming majority of brains in the history of the universe will be Boltzmann brains. According to the ``Copernican principle,'' we should see what a typical observer sees. But we do not, as  most Boltzmann brains find themselves surrounded by thermal equilibrium, not by other intelligent beings on a planet. In fact, if a theory predicts that the great majority of brains are Boltzmann brains, then one should conclude that we are Boltzmann brains. But we are not. We had childhoods. We know because we trust out memories more than any physics theory.

How can any of our serious theories avoid making this incorrect prediction? One possibility is that there will be a Big Crunch, and that the total 4-volume of the universe will be finite and not too large. Then the probability that an unlikely fluctuation ever happens can be small. However, it is expected (e.g., from $\Lambda$-cold-dark-matter, often called the standard model of cosmology) that the late universe will be close to de Sitter space-time and thus has infinite lifetime and infinite 4-volume. We will describe another way out of the problem.\cite{GST15,Tum15}

\section{Concrete version of the Boltzmann brain problem}

It is further expected from $\Lambda$-cold-dark-matter that the state of matter in the late universe will be close (in terms of local observables) to the Bunch-Davies vacuum, a quantum state invariant under the isometries of de Sitter space-time (in particular, a stationary state). The $|\psi|^2$ probability distribution it defines on configuration space gives $>99\%$ weight to thermal equilibrium configurations, but positive probability to brain configurations, in fact $>99\%$ probability to configurations containing brains if 3-space is large enough (in particular if infinite).

Does this mean there are Boltzmann brains in the Bunch-Davies vacuum? Actually, does a stationary state mean that nothing happens? Or does the factual situation visit different configurations over time according to $|\psi|^2$? This is a point at which foundational questions of quantum theory become relevant to cosmology. We need to ask, what is the significance of this particular wave function for reality? 

We will analyze a Bohm-type model of the situation and conclude that there are no Boltzmann brains in the Bunch-Davies state. For comparison, for Everett's many-worlds interpretation of quantum theory, Boddy, Carroll, and Pollack \cite{BCP15} have argued that in a stationary state nothing moves, whereas David Wallace, a leading advocate of the many-worlds view, disagrees (personal communication).

\section{Bohmian mechanics}

Bohmian mechanics \cite{Bohm52,DT} is a version of quantum mechanics with trajectories. For non-relativistic quantum mechanics, it can be defined as follows. The theory takes ``particles'' literally and proposes that electrons are material points moving in 3-space that have a definite position $\vQ_k(t)$ at every time $t$.  There is a wave-particle duality in the sense that there are particles and there is a wave (the usual wave function $\psi_t$ of quantum mechanics on the configuration space $\RRR^{3N}$ of $N$ particles). 

\begin{figure}[h]
\begin{center}
\includegraphics[width=7cm]{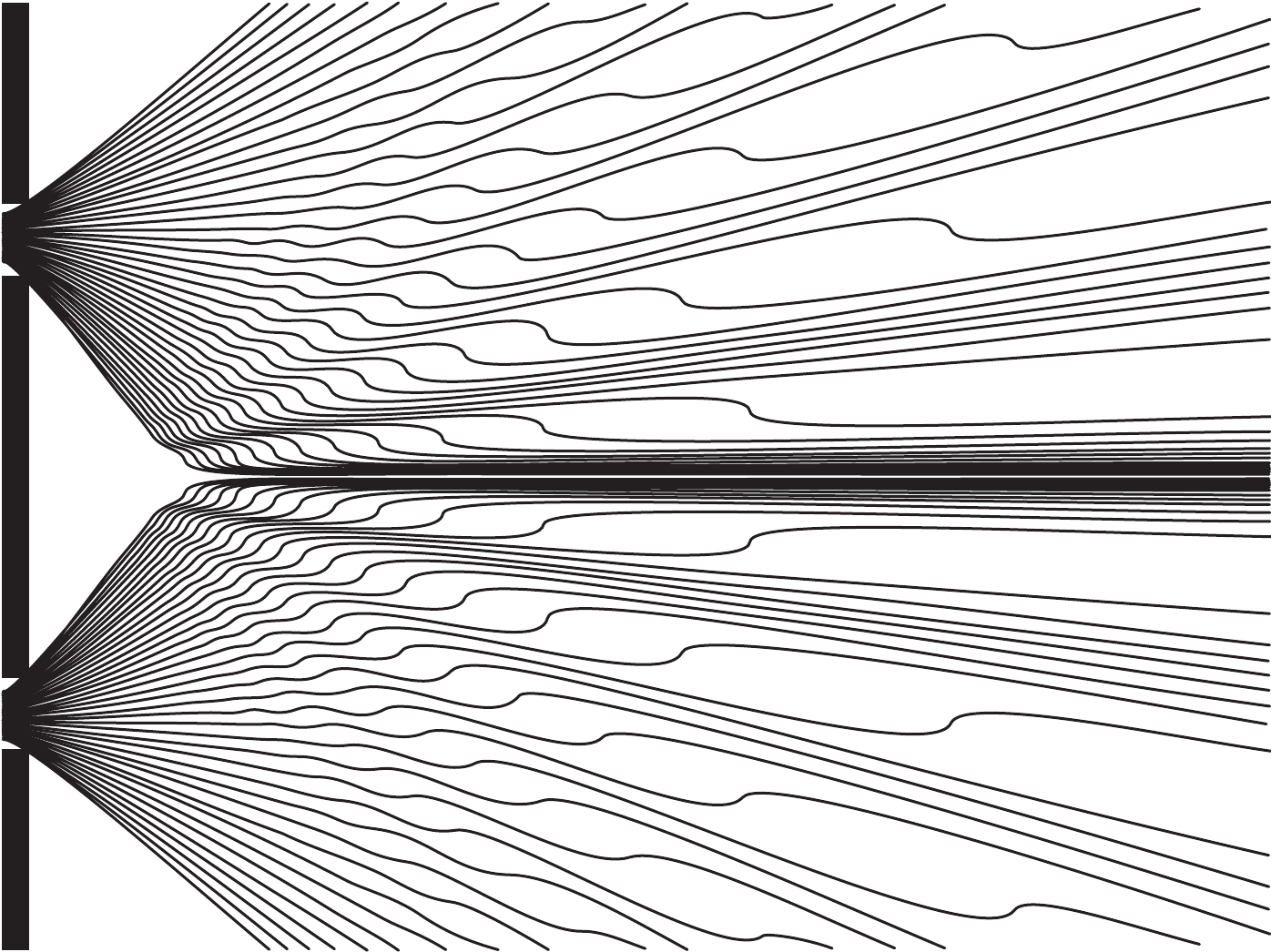}
\end{center}
\caption{Several alternative Bohmian trajectories in a double slit experiment, each coming from the left and passing through one of the slits.
(Reprinted from Ref.~\cite{DT})}
\label{fig1}
\end{figure}

The theory has two dynamical laws: for $\psi_t$ the usual Schr\"odinger equation
\be\label{Schr}
i\hbar\frac{\partial \psi_t}{\partial t} = H\psi\,,
\ee
and for $\vQ_k(t)$ Bohm's equation of motion
\be\label{Bohm}
\frac{d\vQ_k(t)}{dt} = \frac{\hbar}{m_k} \nabla_k \Im \log \psi_t \bigl(\vQ_1(t),\ldots,\vQ_N(t)\bigr)\,.
\ee
Figure~\ref{fig1} shows a sample of Bohmian trajectories after passing through a double slit.
The law of motion \eqref{Bohm} is equivalent to
\be
\frac{dQ}{dt} = \frac{j(Q)}{\rho(Q)}\,,
\ee
where $Q=(\vQ_1,\ldots,\vQ_N)$ is the configuration, $\rho=|\psi|^2$ is the standard probability density, and $j$ is the standard probability current vector field in configuration space. As a consequence of the continuity equation for $\rho$ and $j$, one obtains the \emph{equi\-variance theorem} that if $Q(t=0)$ is random with distribution density $|\psi(t=0)|^2$, then at any time $t$, $Q(t)$ has distribution $|\psi(t)|^2$. As a part of the theory, we assume that the initial configuration is in fact so chosen. 
Figure~\ref{fig1} illustrates that the arrival points on the screen are $|\psi|^2$ distributed.
The central fact about the empirical predictions of Bohmian mechanics is that the inhabitants of a Bohmian world would observe outcomes in agreement with the predictions of quantum mechanics.

\section{How Bohmian mechanics helps against Boltzmann brains}

\subsection{Freezing in Bohmian mechanics}

In Bohmian mechanics, there is the phenomenon of ``freezing,'' whose basic form is as follows.

\begin{thm}
If $\psi$ is a non-degenerate eigenstate of a Hamiltonian that is a non-relativistic Schr\"odinger operator, then the Bohmian configuration does not move. 
\end{thm}

That is because the conjugate of $\psi$ must be another eigenstate with the same eigenvalue, so $\psi$ must be real up to a global phase. As a consequence, according to \eqref{Bohm}, particle velocities vanish. Freezing may be surprising because the momenum distribution, $|\hat\psi(k)|^2$ with $\hat\psi$ the Fourier transform of $\psi$, is not concentrated on the origin. In Bohmian mechanics, momentum corresponds not to the instantaneous velocity but to the asymptotic velocity that the particle would reach if the potential were turned off.\cite{DT}

It follows that, if non-relativistic Bohmian mechanics were true, and if the late universe were in a non-degenerate eigenstate, then the configuration would be frozen. Arguably, the Boltzmann brain problem is absent then, as any such brain, even if it existed, would not be functioning. Turning now to a more realistic scenario, we will see that although freezing as such does not occur in the late universe, asymptotic freezing does.

\subsection{Concrete model of the late universe}

Let us take for granted that the late universe is asymptotically a de Sitter space-time with metric
\be
ds^2 = dt^2 -  e^{2Ht} \delta_{ij} dx^i dx^j\,,
\ee
where $H$ is the Hubble parameter (expansion speed of the universe---not the Hamiltonian!) and $i,j$ are summed over the values $1,2,3$. Actually, this coordinate patch covers only half of de Sitter space-time, but that is alright because we are considering only the late universe. As a model of the matter, we consider a simple quantum field theory with a single Hermitian scalar quantum field $\varphi({\bf x},t)$. The quantum state can then be written as a wave functional $\Psi(\varphi,t)$ on the space of field configurations $\varphi$, which are real-valued functions of ${\bf x}=(x^1,x^2,x^3)$. It will be convenient to rescale time according to $dt=e^{Ht} \, d\eta$, where the new coordinate $\eta$ is called the conformal time. 
As $t$ varies from $-\infty$ to $+\infty$, $\eta$ runs from $-\infty$ only up to $0$; in particular, the long-time limit $t\to \infty$ that is relevant to us corresponds to $\eta \to 0$ from the left. We also rescale the field variable according to $y=e^{Ht}\varphi$. Moreover, it will be useful to express the field configuration in terms of its Fourier modes $y_{\bf k}$ (note that $y_{-{\bf k}}=y^*_{\bf k}$ since $y({\bf x})$ is real).
The Schr\"odinger equation for $\Psi$ in the Schr\"odinger picture then reads\cite{PNSS12,GST15} 
\be\label{kSchr}
i \frac{\partial\Psi}{\partial\eta} = 
\int_{{\mathbb R}^{3+}} \!\! d^3{\bf k} \biggl[ -\frac{\delta^2}{\delta y_{\bf k}^*\delta
y_{\bf k}}+
k^2 \, y_{\bf k}^* \, y_{\bf k}
+ \frac{i}{\eta}\left(\frac{\delta}{\delta y_{\bf k}^*}y_{\bf k}^*+
y_{\bf k}\frac{\delta}{\delta y_{\bf k}}\right)\biggr]\Psi\,,
\ee
where $k=|{\bf k}|$, and $\RRR^{3+}$ means an arbitrary choice of half space (containing exactly one of ${\bf k}$ and $-{\bf k}$ for almost every ${\bf k}$).

We consider a Bohmian model with a field ontology, \cite{HAM95,PNSS12} that is, we assume that there is an actual field configuration $\varphi({\bf x},t)$ (the analog of $Q(t)$), equivalently expressed as $y_{\bf k}(\eta)$, which is ``guided'' by $\Psi$ according to the equation
\be\label{kBohm}
\frac{dy_{\bf k}}{d\eta} = \frac{\delta \Im \log \Psi}{\delta y_{\bf k}^*} -  \frac{1}{\eta}y_{\bf k} \,,
\ee
which is the natural analog in this setting of Bohm's equation of motion \eqref{Bohm}.\cite{PNSS12}

\subsection{Asymptotic freezing in the Bunch-Davies state}

In this representation, the Bunch-Davies state reads
\be
\Psi = \prod_{{\bf k} \in \mathbb{R}^{3+}}
\frac{1}
{\sqrt{2\pi}f} \exp{\left\{-\frac{1}{2f^2}|y_{\bf k}|^2 +  
i \left[\left(\frac{f'}{f}+
\frac{1}{\eta}\right)|y_{\bf k}|^2-\text{phase}(k,\eta) \right]\right\}}
\ee
with $f=f_k(\eta) = \sqrt{1+1/k^2\eta^2}/\sqrt{2k}$.
In particular, the different field modes ${\bf k}$ are disentangled. The phase of $\Psi$ is not constant, so $y_{\bf k}$ is not frozen, but $\Psi$ is simple enough so we can explicitly solve \cite{GST15} the Bohmian equation of motion \eqref{kBohm}: 
\be
y_{\bf k}(\eta) = \tilde{c}_{\bf k} \, f_k(\eta)
\ee
or, expressed in the original variables,
\be\label{asy}
\varphi_{\bf k} (t) = c_{\bf k} \sqrt{1+k^2\exp(-2Ht)/H^2} \,.
\ee

\begin{figure}[h]
\begin{center}
\begin{tikzpicture}
\draw (-0.1,0) -- (2.5,0);
\draw (0,-0.1) -- (0,1.7);
\node at (2.4,-0.2) {$t$};
\draw [ultra thick, domain=0:2.5] plot (\x, {sqrt(0.25+2*exp(-2*\x))} );
\node at (1.5,0.9) {$\varphi_{\bf k}$};
\draw (-0.1,0.5) -- (0,0.5);
\node at (-0.3,0.5) {$c_{\bf k}$}; 
\end{tikzpicture}
\end{center}
\caption{Graph of $\varphi_{\bf k}$ as a function of $t$}
\label{fig2}
\end{figure}
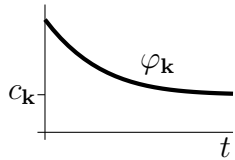

Note that $\lim\limits_{t\to\infty} \varphi_{\bf k}(t)$ exists and is equal to $c_{\bf k}$, which means the Bohmian field is asymptotically freezing. This suggests that the configuration becomes constant for large $t$, precluding a functioning brain. In fact, the situation is a little more subtle because the $x^j$ coordinates are moving apart, so that at any time only the modes with wave lengths large compared to the Hubble distance $1/H$ are frozen. But even before \eqref{asy} becomes constant, its simple behavior is as good as freezing, as it is too simple to support the complex behavior of a functioning brain. Hence, in Bohmian mechanics there are no Boltzmann brains in the Bunch-Davies vacuum.

\subsection{Asymptotic freezing in a generic state}

However, $\Psi$ will not be close to the Bunch-Davies state in Hilbert space. It will look locally similar, but Bohmian mechanics depends nonlocally on the wave function.
This issue is taken care of by the following

\begin{thm}[Ryssens \cite{Rys12} 2012, Tumulka \cite{Tum15} 2015]
For a large class of wave functions $\Psi$ and most initial field configurations, the asymptotic long-time behavior of $\varphi$ is
\be
\varphi_{\bf k}(t) \approx c_{\bf k} \sqrt{1+k^2 \exp(-2Ht)/H^2} \quad \text{for }t>t_0\,,
\ee
where $t_0$ is independent of $\bf k$ (but depends on $\Psi$). In particular, $ \lim\limits_{t\to\infty} \varphi_{\bf k}(t)$ exists.
\end{thm}

\subsection{Idea of proof}

Consider a single mode $\bf k$.
Rescale the field variable, setting $z=\gamma(\eta)^{-1}y$, with the function $\gamma$ to be chosen later. Rescale and phase-transform wave function,
\be
\Phi(z,\eta) = e^{\alpha(\eta)+i\beta(\eta)z^*z} \Psi(\gamma(\eta) \, z,\eta)
\ee
with functions $\alpha$ and $\beta$ to be chosen later.
Also rescale time according to $d\tau = \gamma^{-2} \, d\eta$.
For a suitable (explicitly known \cite{Tum15}) choice of the scaling functions $\alpha,\beta,\gamma$, the evolution of $\Phi$ reduces to a non-relativistic Schr\"odinger equation in a 2d harmonic oscillator potential,
\be
i\frac{\partial \Phi}{\partial \tau} = - \frac{\partial^2 \Phi}{\partial z^*\partial z} + \omega^2 z^*z \Phi
\ee
and the non-relativistic Bohmian equation of motion
\be
\frac{dz}{d\tau} = \frac{\partial \Im\log\Phi(z,\tau)}{\partial z^*}\,.
\ee
These evolutions do not become singular as $\tau\to 0-$ ($\Leftrightarrow \: \eta\to 0- \: \Leftrightarrow\: t\to \infty$).
Thus, $\lim\limits_{\tau \to 0-} z(\tau)$ exists, so, for $\tau$ sufficiently close to 0, 
$z(\tau) \approx \text{const.} \:\Leftrightarrow\: y(\eta) \approx  \gamma(\eta)\times  \text{const.} \:\Leftrightarrow\: \varphi(t) \approx c_{\bf k}\sqrt{1+ k^2\exp(-2Ht)/H^2}$.

\subsection{Upshot}

For almost any $\Psi$, the Bohmian motion becomes very simple as $t\to\infty$. While there is a positive probability for a brain configuration to occur, this subsystem would not function as a brain because it cannot move in a complex way. (In fact, the probability of a brain configuration in a Hubble volume is tiny.)
Thus, Boltzmann brains do not occur in Bohmian mechanics according to this particular model.


\begin{thebibliography}{00}

\bibitem{Alb02} A. Albrecht: 
	Cosmic Inflation and the Arrow of Time. 
	In J.D. Barrow, P.C.W. Davies, C.L. Harper (editors), 
	{\it Science and Ultimate Reality} 
	(Cambridge University Press 2004), 
	\url{http://arxiv.org/abs/astro-ph/0210527}

\bibitem{BCP15} K.K. Boddy, S.M. Carroll, and J. Pollack:
	Why Boltzmann Brains DonÕt Fluctuate Into Existence From the De Sitter Vacuum.
	In K. Chamcham, J. Barrow, J. Silk, and S. Saunders (editors), 
	{\it The Philosophy of Cosmology} 
	(Cambridge University Press 2016),
	\url{http://arxiv.org/abs/1505.02780}

\bibitem{Bohm52} D. Bohm: 
	A Suggested Interpretation of the Quantum Theory in Terms of ``Hidden'' Variables, I and II. 
	\textit{Physical Review} \textbf{85}, 166 (1952)

\bibitem{DT} D.~D\"urr and S.~Teufel:
	{\it Bohmian mechanics} 
	(Springer, Heidelberg, 2009)

\bibitem{GST15} S. Goldstein, W. Struyve, and R. Tumulka:
	The Bohmian Approach to the Problems of Cosmological 
	Quantum Fluctuations.	
	To appear in A.~Ijjas and B.~Loewer (editors), 
	\textit{Guide to the Philosophy of Cosmology} 
	(Oxford University Press, 2019),
	\url{http://arxiv.org/abs/1508.01017}

\bibitem{HAM95} B.J. Hiley and A.H. Aziz Mufti:
	The Ontological Interpretation of Quantum Field Theory Applied in a Cosmological Contex.
	P.\ 141 in M. Ferrero and A. van der Merwe (editors), 
	\textit{Fundamental Theories of Physics} \textbf{73}
	(Kluwer, Dordrecht, 1995)

\bibitem{PNSS12} N. Pinto-Neto, G. Santos, and W. Struyve:
	The quantum-to-classical transition of primordial cosmological perturbations.
	\textit{Physical Review D} \textbf{85}, 083506 (2012),
	\url{http://arxiv.org/abs/1110.1339}

\bibitem{Rys12} W. Ryssens: 
	{\it On the Quantum-to-Classical Transition of Primordial
	Perturbations}, 
	Master thesis, Physics \& Astronomy Department,
	Katholieke Univ. Leuven (2012)

\bibitem{Tum15} R. Tumulka: 
	Long-Time Asymptotics of a Bohmian Scalar Quantum Field 
	in de Sitter Space-Time.
	{\it General Relativity and Gravitation} {\bf 48}, 2 (2016),
	\url{http://arxiv.org/abs/1507.08542}

\end{thebibliography}
\end{document}